\documentstyle[aps, manuscript]{revtex}
\tightenlines
\draft
\begin{document}
\title{Noncompact KK theory of gravity: stochastic
treatment for a nonperturbative inflaton field in a de Sitter expansion}
\author{$^1$Jos\'e Edgar Madriz Aguilar\footnote{
E-mail address: edgar@itzel.ifm.umich.mx}
and $^{2}$Mauricio Bellini\footnote{
E-mail address: mbellini@mdp.edu.ar}}
\address{$^1$Instituto de F\'{\i}sica y Matem\'aticas,
AP: 2-82, (58040) Universidad Michoacana de San Nicol\'as de Hidalgo,
Morelia, Michoac\'an, M\'exico.\\
$^2$Departamento de F\'{\i}sica, Facultad de Ciencias Exactas y Naturales,
Universidad Nacional de Mar del Plata and
Consejo Nacional de Ciencia y Tecnolog\'{\i}a (CONICET),
Funes 3350, (7600) Mar del Plata, Argentina.}

\vskip .5cm
\maketitle
\begin{abstract}
We study a stochastic formalism for a nonperturbative
treatment of the inflaton field
in the framework of a noncompact Kaluza-Klein (KK) theory
during an inflationary (de Sitter) expansion,
without the slow-roll approximation.
\end{abstract}
\vskip .2cm
\noindent
Pacs numbers: 04.20.Jb, 11.10.kk, 98.80.Cq \\
\vskip 1cm
\section{Introduction}

Stochastic inflation
model is one of the very few that solves
almost all of the well-known cosmological problems.
Since the differential microwave radiometer (DMR) mounted on the 
Cosmic Background Explorer satellite (COBE) first detected temperature 
anisotropies in the cosmic background radiation (CBR), we have the 
possibility to directly probe the initial density perturbation.  
The fact that the resulting energy density fluctuations 
(${\delta\rho\over\rho}\approx 10^{-5}$) fit the scaling 
spectrum predicted by the inflation model, suggests that they had 
indeed their origin in the quantum fluctuations of the ``inflaton'' 
scalar field during the inflationary era.  
Although in principle this problem is of a quantal nature, the 
fact that under certain conditions---which are made 
precise in \cite{Pollock,Habib,BCMS}---the inflaton field can be considered 
as classical largely simplifies the approach, by 
allowing a Langevin-like stochastic treatment.
The most
widely accepted approach assume that the inflationary phase is driving by a
quantum scalar field $\varphi $ with a potential $V(\varphi )$. Within this
perspective, the stochastic inflation proposes to describe the dynamics of
this quantum field on the basis of a splitting of $\varphi$ in a homogeneous
and an inhomogeneous components. Usually the homogeneous one $\phi_c(t)$
is interpreted
as a classical field that arises from a coarsed-grained average over a
volume larger than the observable universe, and plays the role of a global
order parameter\cite{goncharov}. All information on scales smaller than this
volume, such as the density fluctuations, is contained in the inhomogeneous
component. Although this theory is widely used and accepted as general
one needs to make the approximation $\left<\rho\right>
\simeq {\dot\phi_c^2 \over 2} +V(\phi_c)$ to can make some calculations
in a linear expansion for the scalar potential $V(\varphi)$ around
its classical background field $\phi_c(t)$\cite{BCMS}.
It was Starobinsky the first one to derive a Fokker-Planck equation 
for the transition probabilities $P(\phi_L,t|\phi'_L,t')$ in 
comoving coordinates\cite{Starobinsky} from the stochastic 
equation for the dynamics of the inflaton field.  
$P(\phi_L,t|\phi'_L,t')$ provides us with statistical information 
about the relative number of ``domains'' (metastable 
vacuum configurations) that having a typical value 
$\phi'_L$ of the coarse-grained inflaton field, evolve in a time 
interval $(t-t')$ towards a new configuration with a typical value 
$\phi_L$.

The main aim of this work consists to make a consistent
coarse-granning treatment
for the scalar field dynamics on cosmological scales
during the inflationary epoch. For
simplicity, as an example, we shall study
a de Sitter expansion for the universe,
but the formalism here developed can be used to study other
more realistic inflationary models.
The strategy consists to starts from a 5D globally flat metric
and an action for a purely kinetic quantum scalar field minimally
coupled to gravity, which define a 5D vacuum state.
The metric we consider can be mapped to a 5D generalized
Friedmann-Roberton-Walker (FRW) on which we make a foliation
by considering the fifth (spatial-like) coordinate as a constant.
As a result of this foliation we obtain an effective 4D FRW metric
and an effective 4D density Lagrangian in which appears an effective term
which depends on the fifth coordinate and is not
kinetic in 4D. This term is identified as a 4D scalar
potential or source, and has an origin purely geometric.
The idea that matter in four dimensions (4D) can be explained from
a 5D Ricci-flat ($R_{AB}=0$) Riemannian manifold is a consequence of the
Campbell's theorem. It says that any analytic $N$-dimensional Riemannian
manifold can be locally embedded in a $(N+1)$-dimensional Ricci-flat
manifold. This is of great importance for establishing the generality of the
proposal that 4D field equations with sources can be locally embedded
in 5D field equations without sources\cite{2,dos}.
The advantage of this propose
is that
provides an exact (nonperturbative) treatment for the 4D
dynamics of the inflaton field (the scalar field) with
back-reaction effects included\cite{PLB2005}. Furthermore, it is
possible to make a
consistent treatment for the effective 4D dynamics of the universe
in other models governed by a single scalar field.
In this paper we consider a general version of the Kaluza - Klein
theory in 5D, where the extra dimension is not assumed to be compactified.
In other words, this means that the cylinder condition in the fifth
coordinate of the original Kaluza - Klein theory is relaxed. From the
mathematical viewpoint, this means that the 5D metric tensor is allowed
to depend explicitly on the fifth coordinate. Without cylindricity, there
is no reason to compactify the fifth dimension, so this approach is properly
called noncompactified.\\

\section{Review of the formalism}

We consider an action
\begin{equation} 
I=-\int d^{4}xd\psi\,\sqrt{\left|\frac{^{(5)}
g}{^{(5)}g_0}\right|} \ \left[\frac{^{(5)}R}{16\pi G}+ ^{(5)}{\cal 
L}(\varphi,\varphi_{,A})\right],
\label{action}
\end{equation}
for a scalar field $\varphi$, which is minimally coupled to gravity.
Here, $|^{(5)}g|=\psi^8 e^{6N}$
is the absolute value of the determinant for the 5D metric tensor with
components $g_{AB}$ ($A,B$ take the values $0,1,2,3,4$) and
$|^{(5)}g_0|=\psi^8_0 e^{6N_0}$ is a constant of dimensionalization 
determined by $|^{(5)}g|$ evaluated at $\psi=\psi_0$ and $N=N_0$.
Furthermore,
$^{(5)}R$ es the 5D Ricci scalar
and $G$ is the gravitational constant.
In this work we shall consider $N_0=0$, so that
$^{(5)}g_0=\psi^8_0$. Here, the index $`` 0 "$ denotes the values
at the end of inflation (i.e.,
when $\ddot b =0$).
With the aim to describe a manifold in apparent vacuum
the Lagrangian density ${\cal L}$ in (\ref{action}) must be only
kinetic in origin
\begin{equation}\label{1'}
^{(5)}{\cal L}(\varphi,\varphi_{,A}) =
\frac{1}{2} g^{AB} \varphi_{,A} \varphi_{,B},
\end{equation}
and the 5D metric should be globally flat\cite{3,new,PLB}
\begin{equation}\label{6}
dS^2 =\psi^2 dN^2 - \psi^2 e^{2N} dr^2 - d\psi^2,
\end{equation}
where $dr^{2}=dx^{2}+dy^{2}+dz^{2}$. The coordinates ($N$,$\vec r$) 
are dimensionless and the fifth coordinate $\psi $ has spatial units.
The metric (\ref{6}) describes a flat 5D manifold 
in apparent vacuum ($G_{AB}=0$), so that
the Ricci scalar must be zero: $^{(5)}R=0$.
We consider a diagonal metric because 
we are dealing only with gravitational effects, which are the 
important ones for the global evolution for the universe\cite{van}.

The equation of motion for the scalar quantum field $\varphi$ is
\begin{equation}\label{df}
\left(2\psi \frac{\partial\psi}{\partial N}+ 3 \psi^2 \right)
\frac{\partial\varphi}{\partial N}
+\psi^2 \frac{\partial^2\varphi}{\partial N^2} - \psi^2 e^{-2N} 
\nabla^2_r\varphi
-4\psi^3 \frac{\partial\varphi}{\partial\psi} - 3\psi^4 \frac{\partial 
N}{
\partial\psi} \frac{\partial\varphi}{\partial\psi} - \psi^4 
\frac{\partial^2
\varphi}{\partial\psi^2} =0,
\end{equation}
where ${\partial N \over \partial\psi}$ and ${\partial\psi \over \partial N}$ are zero because the 
coordinates $(N,\vec{r},\psi)$ are independent.

The eq. (\ref{df}) can be written as
\begin{equation}
\stackrel{\star \star}{\varphi}+3\stackrel{\star}{\varphi}-e^{-2 
N}\nabla_{r}^{2}\varphi -\left[4\psi 
\frac{\partial\varphi}{\partial\psi}+\psi^{2} 
\frac{\partial^{2}\varphi}{\partial\psi^{2}}\right]=0,
\label{ec2}
\end{equation}
where the overstar denotes the derivative with respect to $N$ and 
$\varphi \equiv \varphi(N,\vec{r},\psi)$.
The commutator between $\varphi$ and $\Pi^N = {\partial {\cal L} \over \partial
\varphi_{,N}} = g^{NN} \varphi_{,N}$ is given by
\begin{equation}
\left[\varphi(N,\vec r,\psi), \Pi^N(N,\vec{r'},\psi')\right] =
ig^{NN} \left|\frac{^{(5)} g_0}{^{(5)} g}\right| \  \delta^{(3)}(\vec r-
\vec{r'}) \delta(\psi - \psi'),
\end{equation}
where $\left|{^{(5)} g_0 \over ^{(5)} g}\right|$ is the inverse
of the renormalized volume of the manifold (\ref{6}) and $g^{NN} = \psi^{-2}$.
Hence, the commutator between $\varphi$ and $\stackrel{\star}{\varphi}$ will
be
\begin{equation}\label{commut}
\left[\varphi(N,\vec r,\psi), \stackrel{\star}{\varphi}
(N,\vec{r'},\psi')\right] =
i \left|\frac{^{(5)} g_0}{^{(5)} g}\right| \  \delta^{(3)}(\vec r-
\vec{r'}) \delta(\psi - \psi').
\end{equation}

By means of the transformation $\varphi =\chi 
e^{-3N/2}\left(\psi_{0}\over \psi \right)^{2}$ we obtain the 
5D generalized Klein-Gordon like equation for 
$\chi (N,\vec{r},\psi)$ and the commutator between $\chi$ and
$\stackrel{\star}{\chi}$:
\begin{eqnarray}
&& \stackrel{\star \star}{\chi}-\left[e^{-2 N}\nabla_{r}^{2} 
+\left(\psi^{2} \frac{\partial^2}{\partial\psi^2} +\frac{1}{4}\right) 
\right]\chi =0,  \label{ec3} \\
&& 
\left[\chi(N,\vec r,\psi), \stackrel{\star}{\chi}
(N,\vec{r'},\psi')\right] =
i  \delta^{(3)}(\vec r- \vec{r'}) \delta(\psi - \psi'). \label{commm}
\end{eqnarray}

The redefined field $\chi$ can be written in terms of a Fourier expansion
\begin{eqnarray} \label{ec4}
\chi (N,\vec{r},\psi)&=& \frac{1}{(2\pi)^{3/2}}\int d^{3} k_{r} \int d 
k_{\psi} \left[a_{k_{r}k_{\psi}} e^{i(\vec{k_r} \cdot \vec{r} 
+{k_\psi}\cdot {\psi})}\xi_{k_{r}k_{\psi}}(N,\psi) \right. \nonumber
\\
&+& \left. a_{k_{r}k_{\psi}}^{\dagger} e^{-i(\vec{k_r} \cdot \vec{r} 
+{k_\psi} \cdot {\psi})}\xi_{k_{r}k_{\psi}}^{*}(N,\psi)\right],
\end{eqnarray}
where the asterisk denotes the complex conjugate and 
$(a_{k_{r}k_{\psi}},a_{k_{r}k_{\psi}}^{\dagger})$ are respectively
the annihilation
and creation operators which satisfy the following commutation
expresions
\begin{eqnarray}\label{ec5}
\left[a_{k_{r}k_{\psi}},a_{k_{r}k_{\psi}}^{\dagger}\right]&=&\delta^{(3)
}\left(\vec{k_r}-\vec{k'_r}\right) 
\delta\left(\vec{k_\psi}-\vec{k'_\psi}\right),\\
\label{ec6}
\left[a_{k_{r}k_{\psi}}^{\dagger},a_{k'_{r}k'_{\psi}}^{\dagger}\right]
&=&\left[a_{k_{r}k_{\psi}},a_{k'_{r}k'_{\psi}}\right]=0.
\end{eqnarray}
The expression (\ref{commm}) complies if the modes are renormalized
by the following condition:
\begin{equation} \label{recon}
\xi_{k_{r}k_{\psi}}\left(\stackrel{\star}{\xi}_{k_{r} 
k_{\psi}}\right)^{*} - \left(\xi_{k_{r} k_{\psi}}\right)^{*} 
\stackrel{\star}{\xi}_{k_{r}k_{ \psi }} = i.
\end{equation}
This equation provides the renormalization for the complete set of
solutions on all the spectrum ($k_r,k_{\psi}$).
On the other hand, the dynamics for the
modes $\xi_{k_{r}k_{\psi}}(N,\psi)$ is well described by
the equation
\begin{equation} \label{ec8}
\stackrel{\star \star}{\xi}_{k_{r}k_{\psi}} +k_{r}^{2} e^{-2 N} 
\xi_{k_{r}k_{\psi}} +\psi^{2}\left(k_{\psi}^{2} 
-2ik_{\psi}\frac{\partial}{\partial \psi}-\frac{\partial^{2}}{\partial 
\psi^{2}} -\frac{1}{4\psi^{2}}\right)\xi_{k_{r}k_{\psi}}=0.
\end{equation}
To solve this equation we can make
\begin{equation} \label{ec9}
\xi_{k_{r}k_{\psi}}(N,\psi)=\xi_{k_{r}}^{(1)}(N) 
\xi_{k_{\psi}}^{(2)}(\psi),
\end{equation}
where the dynamics for
$\xi_{k_{r}}^{(1)}(N)$ and $\xi_{k_{\psi}}^{(2)}(\psi)$
are given by the following differential equations
\begin{eqnarray} 
\stackrel{\star \star}{\xi}_{k_{r}}^{(1)} +\left[k_{r}^{2} e^{-2 
N}-\alpha\right] {\xi}_{k_{r}}^{(1)}&=&0,
\label{ec10}\\
\frac{d^{2}\xi_{k_{\psi}}^{(2)}}{d \psi^2} 
+2ik_{\psi}\frac{d\xi_{k_{\psi}}^{(2)}}{d\psi}-\left(k_{\psi}^{2} 
-\frac{(1/4-\alpha)}{\psi^2}\right)\xi_{k_{\psi}}^{(2)}&=&0. \label{ec11}
\end{eqnarray}
being $\alpha$ a dimensionless constant.

The solution of (\ref{ec8}) can be written as\cite{gi27}
\begin{equation} \label{ec17}
\xi_{k_{r}k_{\psi}}(N,\psi)=\frac{i\sqrt{\pi}}{2}e^{-i\vec{k}_{\psi} 
\cdot \vec{\psi}}{\mathcal H}_{\sqrt{\alpha}}^{(2)}[k_{r}e^{-N}]=
e^{-i\vec{k}_{\psi}.\vec\psi} \  \bar\xi_{k_r}(N),
\end{equation}
where ${\mathcal H}_{\nu}^{(1,2)}[x(N)]={\mathcal J}_{\nu}[x(N)]\pm 
i{\mathcal Y}_{\nu}[x(N)]$ are the Hankel functions, ${\mathcal 
J}_{\nu}[x(N)]$ and ${\mathcal Y}_{\nu}[x(N)]$ are the first and second kind 
Bessel functions with $\nu =\sqrt{\alpha}=1/2$ and $x(N)=k_{r} e^{-N}$. 
Furthermore the function $\bar{\xi}_{k_r}(N)$
is given by
\begin{equation}
\bar{\xi}_{k_r}(N)
=\frac{i \sqrt{\pi}}{2} {\cal H}^{(2)}_{1/2}
\left[ k_r e^{-N}\right].
\end{equation}
In other words, $\xi_{k_r k_{\psi}}(N,\psi)=e^{-i \vec{k}_{\psi}.\vec{\psi}}
\bar\xi_{k_r}(N)$, where $\bar\xi_{k_r}(N)$ is a solution of
\begin{equation}\label{mod5D}
\stackrel{\star\star}{\bar\xi}_{k_r} + \left( k^2_r e^{-2N}
-\frac{1}{4}\right)
\bar\xi_{k_r}=0,
\end{equation}
such that the renormalization condition for $\bar\xi_{k_r}(N)$ becomes
\begin{equation}
\bar\xi_{k_{r}}\left(\stackrel{\star}{\bar\xi}_{k_{r} 
}\right)^{*} - \left(\bar\xi_{k_{r}}\right)^{*} 
\stackrel{\star}{\bar\xi}_{k_{r}} = i.
\end{equation}
Note the discrepance of the result (\ref{ec17}) with whole obtained
in\cite{PLB2005}, which relies in the particular choice of the
vacuum. The vacuum here used is general and solves the problem
of the earlier work in the sense that now there is not
dpendence on $\psi$ in $\chi$.

Hence, the field $\chi$ in eq. (\ref{ec4}) can be
rewritten as
\begin{equation}\label{chichi}
\chi(N,\vec r,\psi) =\chi(N,\vec r)
=\frac{1}{(2\pi)^{3/2}} {\Large\int} d^3k_r
{\Large\int} dk_{\psi} \left[ a_{k_r k_{\psi}} e^{i \vec{k_r}.\vec r}
\bar\xi_{k_r}(N) + a^{\dagger}_{k_r k_{\psi}} e^{-i \vec{k_r}.\vec r}
\bar\xi^*_{k_r}(N)\right].
\end{equation}
Finally, the field $\varphi$ is given by
\begin{equation}
\varphi(N,\vec r,\psi) = e^{-\frac{3N}{2}}
\left(\frac{\psi_0}{\psi}\right)^2
\chi(N,\vec r),
\end{equation}
with $\chi(N,\vec r)$ given by eq. (\ref{chichi}).
Note that exponentials $e^{\pm i \vec{k}_{\psi}.\vec{\psi}}$ disappear
in $\chi(N,\vec r)$ and there is not dependence on the fifth coordinate
$\psi$ in this field. This is a very important fact that say us
that the field $\varphi(N, \vec r, \psi)$ propagates only on the
3D spatially isotropic space $r(x,y,z)$, but not on the additional
space-like coordinate $\psi$.

\section{Coarse-granning in 5D}

To study the large scale evolution of the field $\varphi$ on
large 3D spatial scales, we can introduce the field $\chi_L$
\begin{equation}
\chi_L(N,\vec r,\psi) =
\frac{1}{(2\pi)^{3/2}} {\Large\int} d^3k_r
{\Large\int} dk_{\psi} \theta(\epsilon k_0(N) -k_r)
\left[ a_{k_r k_{\psi}} e^{i \vec{k_r}.\vec r}
\bar\xi_{k_r}(N) + c.c.\right],
\end{equation}
where $c.c.$ denotes the complex conjugate of the first term inside
the brackets and $k_0=\sqrt{\alpha} e^N $
is the $N$-dependent wavenumber (related to the 3D spatially isotropic,
homogeneous and flat space $r^2=x^2+y^2+z^2$), which
separates the long ($k^2_r \ll k^2_0$) and short ($k^2_r \gg k^2_0$)
sectors. Modes with $k_r/k_0 <\epsilon$ are referred to as
outside the horizon.

If the short wavelenght modes are described with the field $\chi_S$
\begin{equation}
\chi_S(N,\vec r,\psi) = \frac{1}{(2\pi)^{3/2}} {\Large\int} d^3k_r
{\Large\int} dk_{\psi} \theta(k_r-\epsilon k_0(N))
\left[ a_{k_r k_{\psi}} e^{i \vec{k_r}.\vec r}
\bar\xi_{k_r}(N) + c.c.\right],
\end{equation}
such that $\chi=\chi_L + \chi_S$, hence
the equation of motion for $\chi_L$
will be approximately
\begin{equation}\label{sto1}
\stackrel{\star\star}{\chi}_L - \left(\frac{k_0(N)}{a}\right)^2 \chi_L =
\epsilon \left[\stackrel{\star\star}{k_0} \eta(N,\vec r,\psi) +
\stackrel{\star}{k_0} \kappa(N,\vec r,\psi) + 2 \stackrel{\star}{k_0}
\gamma(N,\vec r,\psi)\right],
\end{equation}
where the stochastic operators $\eta$, $\kappa$ and $\gamma$ are given
respectively by
\begin{eqnarray}
\eta(N,\vec r,\psi) &=&
\frac{1}{(2\pi)^{3/2}} {\Large\int} d^3k_r
{\Large\int} dk_{\psi} \  \delta(\epsilon k_0(N) -k_r)
\left[ a_{k_r k_{\psi}} e^{i \vec{k_r}.\vec r}
\bar\xi_{k_r}(N) + c.c.\right], \\
\kappa(N,\vec r,\psi) & =&
\frac{1}{(2\pi)^{3/2}} {\Large\int} d^3k_r
{\Large\int} dk_{\psi} \  \stackrel{\star}{\delta}(\epsilon k_0(N) -k_r)
\left[ a_{k_r k_{\psi}} e^{i \vec{k_r}.\vec r}
\bar\xi_{k_r}(N) + c.c.\right],\\
\gamma(N,\vec r,\psi) & = &
\frac{1}{(2\pi)^{3/2}} {\Large\int} d^3k_r
{\Large\int} dk_{\psi} \  \delta(\epsilon k_0(N) -k_r)
\left[ a_{k_r k_{\psi}} e^{i \vec{k_r}.\vec r}
\stackrel{\star}{\bar\xi}_{k_r}(N) + c.c.\right].
\end{eqnarray}
The equation (\ref{sto1}) can be rewritten as
\begin{equation}\label{sto2}
\stackrel{\star\star}{\chi}_L - \alpha \chi_L =
\epsilon \left[\frac{d}{dN}\left(\stackrel{\star}{k_0}
\eta(N,\vec r,\psi)\right) +
\stackrel{\star}{k_0} \gamma(N,\vec r,\psi)\right].
\end{equation}
This is a
second order stochastic equation that can be written as two
first order stochastic ones by introducing
the auxiliar field
$u= \stackrel{\star}{\chi_L} - \epsilon \stackrel{\star}{k_0}\eta$
\begin{eqnarray}
\stackrel{\star}{u} &=& \alpha \chi_L +
\epsilon \stackrel{\star}{k_0} \gamma, \label{uno} \\
\stackrel{\star}{\chi}_L &=&
u + \epsilon \stackrel{\star}{k_0} \eta.\label{dos}
\end{eqnarray}
In the system (\ref{uno}), (\ref{dos}) the role of the
noise $\gamma$ can be minimized 
if $\left(\stackrel{\star}{k_0}\right)^2
\left<\gamma^2\right> \ll \left(\stackrel{\star\star}{k_0}\right)^2
\left<\eta^2\right>$, which holds
if the following condition holds
\begin{equation}\label{condition}
\frac{\stackrel{\star}{\bar\xi_{k_r}} \stackrel{\star}{\bar\xi}^*_{k_r}}{
\bar\xi_{k_r} \bar\xi^*_{k_r}} \ll 1.
\end{equation}
In such case the
system (\ref{uno}), (\ref{dos}) can be approximated to
\begin{eqnarray}
\stackrel{\star}{u} &=& \alpha \chi_L , \label{uno1} \\
\stackrel{\star}{\chi}_L &=& u + \epsilon \stackrel{\star}{k_0}
\eta.\label{dos2}
\end{eqnarray}
This system represents two Langevin equations with a noise
$\eta$ which is gaussian and white in nature
\begin{eqnarray}
&& \left< \eta\right> = 0, \\
&& \left< \eta^2\right> = \frac{\epsilon \left(k_0\right)^2}{2\pi^2
\stackrel{\star}{k_0}} {\Large\int} dk_{\psi} \  \bar\xi_{\epsilon k_0}
\bar\xi^*_{\epsilon k_0} \  \delta(N-N').
\end{eqnarray}
The equation that describes the dynamics of the transition probability
$P\left[\chi^{(0)}_L, u^{(0)}|\chi_L, u\right]$
from a configuration ($\chi^{(0)}_L, u^{(0)}$) to ($\chi_L, u$) is
a Fokker-Planck one
\begin{equation}
\frac{\partial P}{\partial N} = - u \frac{\partial P}{\partial\chi_L}
-\alpha \chi_L \frac{\partial P}{\partial u}
+ \frac{1}{2} D_{11} \frac{\partial^2 P}{\partial \chi_L^2},
\end{equation}
where $D_{11}={1 \over 2} \left(\epsilon \stackrel{\star}{k_0}\right)^2
\left[\int dN \left<\eta^2\right>\right]$
is the diffusion coefficient related to the variable $\chi_L$
due to the stochastic action of the noise $\eta$. Explicitely
\begin{equation}
D_{11} = \frac{\epsilon^3 \left(k_0\right)^2}{4\pi^2}
\stackrel{\star}{k_0} {\Large\int} dk_{\psi} \  \bar\xi_{\epsilon k_0}
\bar\xi^*_{\epsilon k_0},
\end{equation}
which is divergent.

\section{Ponce de Leon metric and 4D de Sitter expansion}

In order to describe
the metric (\ref{6}) in physical coordinates we can make the
following transformations:
\begin{equation}\label{map}
t = \psi_0 N, \qquad R= \psi_0 r, \qquad \psi= \psi,
\end{equation}
such that we obtain the 5D metric
\begin{equation}\label{m1}
dS^2 = \left(\frac{\psi}{\psi_0}\right)^2 \left[dt^2
- e^{2t/\psi_0} dR^2\right]- d\psi^2,
\end{equation}
where $t$ is the cosmic time and $R^2=X^2+Y^2+Z^2$.
This metric is the Ponce de Leon one\cite{PDL}, and describes a
3D spatially flat, isotropic and homogeneous extended (to 5D)
FRW metric in a de Sitter expansion\cite{librowesson}.

To study the de Sitter evolution of the universe
on the 4D spacetime we
can take a foliation $\psi=\psi_0$ in the metric
(\ref{m1}), such that the effective 4D metric
results
\begin{equation}\label{m2}
dS^2 \rightarrow ds^2 = dt^2 - e^{2t/\psi_0} dR^2,
\end{equation}
which describes 4D globally isotropic and homogeneous expansion of
a 3D spatially flat, isotropic and homogeneous universe that expands
with a Hubble parameter $H=1/\psi_0$ (in our case a constant)
and has a 4D scalar curvature
$^{(4)}{\cal R} = 6(\dot H + 2 H^2)$. Note that in this particular
case the Hubble parameter is constant so that $\dot H = 0$.

The 4D energy 
density $\rho$ and the pressure ${\rm p}$ are
\begin{eqnarray}
&& 8 \pi G \left<\rho\right> = \frac{3}{\psi^2_0},\\
&& 8\pi G \left<{\rm p}\right> = -\frac{3}{\psi^2_0},
\end{eqnarray}
where $G=M^{-2}_p$ is the gravitational constant and
$M_p=1.2 \  10^{19} \  GeV$ is the Planckian mass.
Furthermore, the universe describes a vacuum
equation of state: ${\rm p} = - \rho$,
such that
\begin{equation}
\left<\rho\right> = \left<\frac{\dot\varphi^2}{2} + \frac{a^2_0}{2a^2}
\left(\vec\nabla \varphi\right)^2 + V(\varphi)\right>,
\end{equation}
where the brackets denote the 4D expectation vacuum
and the cosmological constant $\Lambda $ gives the vacuum energy density
$\left<\rho\right> = {\Lambda \over 8\pi G}$. Thus, $\Lambda $ is
related with the fifth coordinate by means of $\Lambda = 3/\psi^2_0$\cite{PDL}.
Furthermore, the 4D Lagrangian is given by
\begin{equation}
^{(4)}{\cal L}(\varphi,\varphi_{,\mu}) =
-\sqrt{\left|\frac{^{(4)}g}{^{(4)}g_0}\right|} \left[
\frac{1}{2} g^{\mu\nu} \varphi_{,\mu}\varphi_{,\nu} + V(\varphi)\right],
\end{equation}
where the effective potential for the 4D FRW metric\cite{MB}, is
\begin{equation}\label{pot}
V(\varphi) = -\left.\frac{1}{2} g^{\psi\psi} \varphi_{,\psi}\varphi_{,\psi}
\right|_{\psi=\psi_0} =
\frac{1}{2} \left.
\left(\frac{\partial\varphi}{\partial\psi}\right)^2\right|_{\psi=\psi_0}.
\end{equation}
In our case this potential takes the form
\begin{equation}\label{pot2}
V(\varphi) = \frac{2}{\psi_0^2} \  \varphi^2(t,\vec R,\psi_0),
\end{equation}
where $k_{\psi_0}$ is the wavenumber for $\psi=\psi_0$.
Notice this potential has a geometrical origin and assume different
representations in different frames. In our case the observer is in a
frame $U^{\psi}=0$, because we are taking a foliation $\psi=\psi_0$
on the 5D metric (\ref{m1}).
Furthermore, the effective 4D motion equation for $\varphi$ is
\begin{equation}
\ddot\varphi + \frac{3}{\psi_0} \dot\varphi -
e^{-2t/\psi_0} \nabla^2_R\varphi - \left.\left[
4\frac{\psi}{\psi^2_0} \frac{\partial\varphi}{\partial\psi} +
\frac{\psi^2}{\psi^2_0} \frac{\partial^2\varphi}{\partial\psi^2}\right]
\right|_{\psi=\psi_0} =0,
\end{equation}
which means that the effective derivative (with respect to $\varphi$)
for the potential, is
\begin{equation}
\left.V'(\varphi)\right|_{\psi=\psi_0} =
\frac{2}{\psi^2_0} \  \varphi(\vec R,t,\psi_0).
\end{equation}

Now we can make the following transformation:
\begin{equation}
\varphi(\vec R, t) = e^{-\frac{3t}{2\psi_0}} \chi (\vec R, t).
\end{equation}
Note that now $\varphi\equiv \varphi(\vec R=\psi_0 r,t=\psi_0 N,
\psi=\psi_0)=
e^{-3t/(2\psi_0)} \chi(\vec R, t)$, where [see eq. (\ref{chichi})]
$\chi(t,\vec R) = \chi(t=\psi_0 N, \vec R=\psi_0 r, \psi=\psi_0)$:
\begin{equation}
\chi(\vec R, t) = \frac{1}{(2\pi)^{3/2}}
{\Large\int} d^3 k_R {\Large\int} dk_{\psi} \left[
a_{k_R k_{\psi}} e^{i \vec{k_R}.\vec{ R}}
\bar\xi_{k_R}(t) + c.c.
\right] \  \delta\left( k_{\psi} - k_{\psi_0}\right).
\end{equation}
Hence, we obtain the following 4D Klein-Gordon equation
for $\chi $
\begin{equation}   \label{mod}
\ddot\chi - \left[e^{-\frac{2t}{\psi_0}} \nabla^2_R +\frac{1}{4\psi^2_0}
\right]\chi =0.
\end{equation}
The equation of motion for the time dependent
modes $\bar\xi_{k_R}(t)$ is
\begin{equation}\label{mot}
\ddot{\bar{\xi}}_{k_R} + \left[ k^2_R e^{-\frac{2t}{\psi_0}} 
-\frac{1}{4\psi^2_0}\right] \bar{\xi}_{k_R } =0.
\end{equation}
It is important to notice that eq. (\ref{mot}) is exactly the equation
(\ref{mod5D}) with the variables transformation (\ref{map}),
on the hypersurface $\psi=\psi_0$.

\subsection{4D stochastic dynamics for $\chi_L$
in a de Sitter expansion}

As was made in 5D, we can define the fields $\chi_L(t,\vec R)$
and $\chi_S(t,\vec R)$, which describe respectively the
long and short wavelenght sectors of the field $\chi$.
\begin{eqnarray}
\chi_L(t, \vec R) &=& \frac{1}{(2\pi)^{3/2}} {\Large\int} d^3k_R
{\Large\int} \  \theta(\epsilon k_0(t) - k_R) \left[
a_{k_R k_{\psi}} e^{i \vec{k_R}.\vec{R}} \bar\xi_{k_R}(t) + c.c.\right]
\delta(k_{\psi} - k_{\psi_0}), \\
\chi_S(t, \vec R) &=& \frac{1}{(2\pi)^{3/2}} {\Large\int} d^3k_R
{\Large\int} \  \theta(k_R- \epsilon k_0(t)) \left[
a_{k_R k_{\psi}} e^{i \vec{k_R}.\vec{R}} \bar\xi_{k_R}(t) + c.c.\right]
\delta(k_{\psi} - k_{\psi_0}),
\end{eqnarray}
where $k_0(t) = {1\over 2\psi_0} e^{t/\psi_0}$. The field
that describes the dynamics of $\chi$ on the infrared sector
($k^2_R \ll k^2_0$) is $\chi_L$. Its dynamics obeys
the Kramers-like stochastic equation
\begin{equation}
\ddot\chi_L - \frac{e^{-\frac{2t}{\psi_0}}}{4\psi^2_0} \chi_L =
\epsilon \left[ \frac{d}{dt}\left(\dot{k}_0 \eta(t,\vec R) \right)+
\dot{k}_0 \gamma(t,\vec R)\right],
\end{equation}
where the stochastic operators $\eta$, $\kappa$ and $\gamma$ are
\begin{eqnarray}
\eta &=& \frac{1}{(2\pi)^{3/2}} {\Large\int} k^3 k_R \  \delta(\epsilon k_0-k_R)
\left[a_{k_R k_{\psi_0}} e^{i \vec{k_R}.\vec{R}} \bar\xi_{k_R}(t)
+ c.c.\right],\\
\gamma &=& \frac{1}{(2\pi)^{3/2}}
{\Large\int} k^3 k_R \  \delta(\epsilon k_0-k_R)
\left[a_{k_R k_{\psi_0}} e^{i \vec{k_R}.\vec{R}} \dot{\bar\xi}_{k_R}(t)
+ c.c.\right].
\end{eqnarray}
This second order stochastic equation can be rewritten as two
Langevin stochastic equations
\begin{eqnarray}
&& \dot u = \frac{e^{-\frac{2t}{\psi_0}}}{4\psi^2_0} \chi_L
+ \epsilon \dot k_0 \gamma,\\
&& \dot\chi_L = u + \epsilon \dot k_0 \eta,
\end{eqnarray}
where $u=\dot\chi_L - \epsilon \dot{k}_0 \gamma$.
The condition to can neglect the noise $\gamma$ with respect to
$\eta$, now holds
\begin{equation}\label{ju}
\frac{\dot{\bar\xi}_{k_R} \dot{\bar\xi}^*_{k_R}}{
\bar\xi_{k_R} \bar\xi^*_{k_R}} \ll \frac{\left(\ddot{k}_0\right)^2}{
\left( \dot{k}_0\right)^2},
\end{equation}
on super Hubble scales. Notice this result is exactly the same in
eq. (\ref{condition}), with the transformation (\ref{map}).
For a de Sitter expansion eq. (\ref{ju}) becomes $k_R/k_0 < \epsilon \ll 1$.
It means that the noise $\gamma$ can be neglected on scales
$k_R \ll {e^{t/\psi_0} \over \psi_0}$ (i.e., on super Hubble scales).
The Fokker-Planck equation for the transition probability
$P(\chi^{(0)}_L,u^{(0)}|\chi_L,u)$ is
\begin{equation}
\frac{\partial P}{\partial t} = - u \frac{\partial P}{\partial \chi_L}
-\frac{e^{-\frac{-2t}{\psi_0}}}{4\psi^2_0} \chi_L \frac{\partial P}{\partial
u} + D_{11}(t) \frac{\partial^2 P}{\partial\chi^2_L},
\end{equation}
where $D_{11}(t) = {\epsilon^3\dot k_0 k^2_0\over
4\pi^2 }\left|\bar\xi_{\epsilon k_0}\right|^2$.
Hence, the equation of motion for $\left< \chi^2_L\right> =
\int d\chi_L du \chi^2_L P(\chi_L,u)$ is
\begin{equation}\label{mot1}
\frac{d}{dt}\left<\chi^2_L\right> = D_{11}(t) \simeq
\frac{\epsilon^2 e^{\frac{3 t}{\psi_0}}}{32 \pi^2 \psi^3_0}.
\end{equation}
In order to return to the original field $\varphi_L =
e^{-{3t\over2\psi_0}} \chi_L$ the equation (\ref{mot1}) can be
rewritten as
\begin{equation}
\frac{d}{dt}\left<\varphi^2_L\right> = -\frac{3}{\psi_0} \left<\varphi^2_L
\right> + \frac{\epsilon^2 }{32 \pi^2 \psi^3_0},
\end{equation}
which has the following solution
\begin{equation}
\left<\varphi^2_L\right> = \frac{\epsilon^2}{96 \pi^2 \psi^2_0}
\left[1+ C \   e^{-\frac{3t}{\psi_0}}\right],
\end{equation}
where $C$ is a constant of integration.
When ${3t\over \psi_0} \ll 1$, one obtains [for $\epsilon^2 (1+C)=24$]
\begin{equation}
\left<\varphi^2_L\right>_{\frac{3t}{\psi_0} \ll 1} \simeq
\frac{1}{4 \pi^2 \psi^2_0}\left[1-\frac{3t}{\psi_0}\right] \equiv
\frac{H^2}{4 \pi^2}\left[1-3 Ht \right].
\end{equation}
However, after the end of inflation, when ${3t \over \psi_0} \gg 1$,
it becomes
\begin{equation}
\left< \varphi^2_L \right>_{\frac{3t}{\psi_0} \gg 1} \simeq
\frac{\epsilon^2 H^2}{96 \pi^2},
\end{equation}
which is valid only for $\epsilon^2 \ll 1$.

In order to understand better this result in the context
of the inflaton field fluctuations $\phi(\vec R, t)$, we
can make the following semiclassical approach:
\begin{equation}
\varphi(\vec R, t) = \left< \varphi(\vec R,t)\right> + \phi(\vec R,t),
\end{equation}
where $\left<\varphi\right> = \phi_c(t)$ and
$\left<\phi\right>=0$.
With this representation one obtains
\begin{equation}
\left<\varphi^2\right> = \phi^2_c + \left<\phi^2\right>,
\end{equation}
where $\phi_c(t)$ is the solution of the equation
\begin{equation}\label{phi_c}
\ddot\phi_c + \frac{3}{\psi_0} \dot\phi_c + \frac{2}{\psi^2_0} \phi_c =0.
\end{equation}
The general solution of the differential equation
(\ref{phi_c}) is
\begin{equation}
\phi_c(t) = \phi^{(0)}_c 
e^{-\frac{t}{\psi_0}} \left(1+
C_1 e^{-\frac{t}{\psi_0}}\right),
\end{equation}
where $C_1$ is a constant of integration and
$\phi^{(0)}_c=\phi_c(t=t_i)$, being $t_i$ the time when
inflation starts. Note that after inflation ends
$\phi_c(t\rightarrow \infty) \rightarrow 0$. Hence,
after inflation one obtains the following result:
\begin{equation}
\left<\varphi^2\right>_{{3t\over\psi_0}\gg 1} \simeq
\left<\phi^2\right>_{{3t\over\psi_0}\gg 1},
\end{equation}
which means that for ${3t\over\psi_0} \gg 1$ the following
approximation is fulfilled:
\begin{equation}
\left<\phi^2_L\right>_{{3t\over\psi_0} \gg 1} \simeq
\left<\varphi^2_L\right>_{{3t\over\psi_0} \gg 1} \simeq 
\frac{\epsilon^2 H^2}{96 \pi^2}.
\end{equation}
This is an important result which say us that
the expectation value for the
second momenta of the field $\varphi_L$ at the end of inflation
is approximately given by the expectation value for the
inflaton field fluctuations on cosmological scales (for $\epsilon^2 \ll 1$).

We can estimate the amplitude of density energy fluctuations on cosmological
scales
\begin{equation}
\left.\frac{\delta\rho}{\rho}\right|_{{\rm end}} \simeq
\frac{\left<V'(\varphi)\right>}{\left<V(\varphi)\right>} \  \left<
\phi^2_L\right>^{1/2} \simeq
\frac{\phi_c}{\left<\phi^2_L\right>} \left<\phi^2_L\right>^{1/2}.
\end{equation}
In order to obtain $\left.{\delta\rho \over\rho}\right|_{{\rm end}} \simeq
10^{-5}$, the value of $\phi_c$ at this moment should be (taking $\epsilon=
10^{-3}$)
\begin{equation}
\left.\phi_c\right|_{{\rm end}} \simeq
\frac{0.66 \  10^{-10}}{\psi_0} = 0.66 \  10^{-10} \  H.
\end{equation}
Finally, we can estimate the initial value for $\phi_c$: $\phi^{(0)}_c$.
If we consider $t_{end} \simeq 10^{10} \  G^{1/2}$, we obtain
\begin{equation}
\phi^{(0)}_c \lesssim  M_p,
\end{equation}
for $H \lesssim 10^{-10} \  M_p$ (or $\psi_0 \gtrsim 10^{10} \  G^{1/2}$).
Hence, the value of $\phi_c$ when inflation
starts assumes sub Planckian values.

\section{Conclusions}

In this work we have developed a stochastic treatment for the
effective 4D inflaton field from a KK theory of gravity
without the hypothesis of a slow - roll regime.
In this framework the long - wavelength modes
of the inflaton field reduces to a quantum system subject to a quantum
noise originated by the short - wavelength sector.
In this approach, the effective 4D potential is quadratic in $\varphi $
and has a geometrical origin. Hence, as in STM theory of gravity
4D source terms are induced from a 5D vacuum and the
fifth dimension (here a space-like dimension) is noncompact.
In our theory the 5D
vacuum is represented by a 5D globally flat metric (related
with a $R_{AB}=0$ manifold) and a purely
kinetic density Langrangian for a quantum scalar field minimally coupled
to gravity.
Since the treatment for the scalar field is nonperturbative
a very important feature of this formalism is that
back reaction effects are included in the calculations in a consistent
manner.
An important result here obtained is that
the expectation value for the
second momenta for the field $\varphi_L$ at the end of inflation
is approximately given by the expectation value for the
inflaton field fluctuations on cosmological scales, being
both determinated by the value of the fifth coordinate on
which we take the foliation: $\psi=\psi_0$.
Furthermore, the initial value for the background (and spatially homogeneous)
inflaton field take sub Planckian values. This fact is very important
because solves the problem of initial conditions in other treatments
of chaotic inflation, in which $\phi^{(0)}_c$ assumes trans Planckian
values. \\

\vskip .2cm
\centerline{\bf{Acknowledgements}}
\vskip .2cm
\noindent
JEMA acknowledges CONACyT and IFM of UMSNH (M\'exico)
for financial support.
MB acknowledges CONICET and UNMdP (Argentina)
for financial support.\\

\end{document}